\documentclass{mn2e}
\newcommand{\AU}{\mbox{\footnotesize AU}}

\usepackage{epsfig}
\usepackage{color}

\title[Upper Palaeolithic extinctions and the Taurid Complex]{Palaeolithic extinctions and the Taurid Complex}
\author[W.M. Napier]
{
W.M. Napier$^1$\thanks{E-mail: napierwm@cardiff.ac.uk(WMN)}\\
$^{1}$Cardiff Centre for Astrobiology, Cardiff University, 2 North
Road, Cardiff CF10 3DY, UK \\
}

\begin{document}

\date{Accepted 2010 February 21.  Received 2010 February 20; in original form 2010 January 23}

\maketitle

\label{firstpage}

\begin{abstract}
Intersection with the debris of a large (50-100~km) short-period comet during the Upper Palaeolithic provides a satisfactory explanation for the catastrophe of celestial origin which has been postulated to have occurred around 12900 BP, and which presaged a return to ice age conditions of duration $\sim$1300 years. The Taurid Complex appears to be the debris of this erstwhile comet; it includes at least 19 of the brightest near-Earth objects. Sub-kilometre bodies in meteor streams may present the greatest regional impact hazard on timescales of human concern.
\end{abstract}


\section{Introduction}

The sudden onset of the Younger Dryas cooling 12,900 years ago was marked by intense wildfires over North America, major disruption of human culture, and the rapid extinction of 35 genera of North American mammals (Faith \& Surovell 2009). A thin carbon-rich black layer of this age has been identified at many sites across North America (Haynes 2008), coincident in age with the Younger Dryas boundary.  

Recently, several geochemical markers at this layer have been presented which seem to indicate that a major extraterrestrial event was involved in these events (Firestone et al. 2007; Firestone 2009). The markers include: 

\textit{Nanodiamonds}, found in abundance at fifteen Younger Dryas Boundary (YDB) sites across North America and north-west Europe, covering a quarter of the Earth's surface. They are mostly found embedded within particles of melted plant resins, confined to the thin black layer, none having yet been found in other strata dating from 55,000 BC to the present. All three diamond allotropes are present in the boundary sediments including lonsdaleite (hexagonal nanodiamonds), which is shock-synthesised and found on Earth only in association with ET impacts or inside meteorites (Kennett et al. 2009). The nanodiamonds and lonsdaleite found in meteorites were already present before atmospheric entry (Clarke et al. 1981). 

\textit{Soot, microspherules and magnetic grains} found at high concentrations, the microspherules having trace element abundances comparable to lunar KREEP but not to any other terrestrial or observed meteoritic source (Firestone 2009). The only previously known co-occurrence of soot, nanodiamonds and rapid extinction is at the Cretaceous-Tertiary boundary layer.

To cause destruction on a continental scale by a single impact, the energy required is $\sim$10$^7$~Mt, corresponding to a 4~km wide comet (Toon et al. 1997). Firestone et al. (2007) proposed that such a comet broke up on atmospheric entry and struck the 2-km thick Laurentide ice sheet.  They speculated that four deep holes in the Great Lakes might be craters produced by this event. This short-term catastrophe was followed by a sudden return to ice age conditions on Earth which lasted for over a millennium, the cause of which is a matter for speculation (loc. cit.; Franz\'{e}n \& Cropp 2007). An alternative proposal to explain the data is that the Earth encountered a rare swarm of carbonaceous chondrites or comets, yielding multiple airbursts and possible surface impacts (Kennett, Kennett, West et al. 2009). 
 
The nature of the event at the Younger Dryas Boundary remains controversial. For example it has been argued that the mammoths were in decline for about a millennium before the final extinctions (Gill, Williams, Jackson et al. 2009). Nevertheless it seems difficult to account for the geophysical markers -- nanodiamonds, exotic sediment composition, evidence of a sudden, continent-wide fiery catastrophe etc -- without the occurrence of some extraordinary extraterrestrial event. 

Problems have however been raised from the astronomical perspective. It is difficult to reconcile a major astronomical catastrophe of continental dimensions as recently as 12.9~ka ago with the known population of near-Earth objects (NEOs) currently being revealed by comet and asteroid search programmes such as the LINEAR and Catalina surveys.  Thus it has been estimated that a 10 megaton impact may occur anywhere on Earth with a mean recurrence time 2000--3000 years (Stuart \& Binzel 2004). An impact of energy 1000~Mt or more is expected with recurrence time $\sim$60,000~yr (loc. cit. and Morbidelli, Jedicke, Bottke et al. 2002). While this would be immensely damaging, the effect would be regional rather than continental or global. On conventional reckoning, the impact of a 4~km asteroid as recently as 13,000 years ago is a one in a thousand event, and if an active comet, the probability is a few in a million (Harris 2008). It has also been pointed out that, to ignite fires in southern Arizona by thermal radiation from a fireball over the Great Lakes, the projectile would have had to be 100 km across, and all non-microbial life on Earth would have been wiped out (Melosh 2009). The geophysical evidence for an astronomical catastrophe at this boundary has been extended to Venezuela (Mahaney et al. 2010), which exacerbates that difficulty.

However, a potential deficiency of impact assessments based on current Spaceguard data is that they assume statistical completeness. To estimate impact risks at say the $\sim$10$^{-4}$ per annum level by extrapolating from $\sim$10~yr of Spaceguard surveys is analogous to forecasting three years' weather by extrapolating one day's observation. A cascade of comet disintegration, for example, could be a significant hazard on such timescales (cf the 1994 impacts on Jupiter following the disintegration of D/1993 F2 Shoemaker-Levy~9). 

It has also been objected that, if the Younger Dryas event was due to an encounter with a swarm of debris 13,000 years ago, the material would still be visible in the inner solar system at the present time. Meteoroid data, being largely tracers of comet disintegration, do indeed have the potential to look into the history of the cometary environment for tens of thousands of years, and in this paper we discuss such evidence. We first, in Section 2, summarise evidence which indicates that a large, short-period comet entered the inner planetary system a few 10$^4$~yr ago and has since then been undergoing a hierarchy of disintegrations, yielding the Taurid Complex. We show that at least 19 of the largest NEOs have orbits significantly close to that of Comet Encke. In Section 3 we numerically model the disintegration history of the progenitor comet. This reveals that encounters with dense swarms of material, sufficient to produce a 12.9~ka cosmic event, are indeed reasonable expectations of recent Earth history. Section 4 gives a broadbrush description of the likely consequences of an encounter with such a swarm. There is satisfactory agreement with the main physical features at this boundary.

The occurrence of a major celestial catastrophe over continental dimensions as recently as 12900 years ago, which is quite unexpected from Spaceguard data alone, may carry implications for assessments of the impact hazard currently faced by civilization (cf Napier \& Asher 2009). 

\section{A large Palaeolithic short-period comet}

\subsection{ The zodiacal cloud and the Taurid meteors}

The complex nature of the Arietid-Taurid meteor streams has long been recognized, with for example Denning (1928) identifying 13 separate sub-streams within the system. The Complex encompasses comet 2P/Encke with orbital period 3.3~yr. Differential orbital precession arising from Jovian perturbations yields two main night-time branches, and two daytime branches. From the orbital dispersion of the material, Steel \& Asher (1996) derive an age 20--30~kyr for the system (cf Babadzhanov et al. 1990).

This broad complex, which comprises at least half the mass of the zodiacal cloud, appears not to be in equilibrium and to be replenished sporadically (Napier 2001).
Observations of both visual (Stohl 1986) and radar `sporadic' meteors (Campbell-Brown 2008) indicate that they are not in fact sporadic but are rather concentrated in a small number of primary sources, two of which -- the helion and and antihelion sources -- appear to derive from Comet 2P/Encke or its parent. Wiegert, Vaubaillon \& Campbell-Brown (2009a,b), modelling this sporadic meteor complex, find that this comet dominates the helion and antihelion sources by a factor of about ten. They consider that the sporadic population is dominated by a small number of comets (especially Encke and 55P/Tempel-Tuttle) with high-transfer efficiencies to Earth, rather than the comet population as a whole. The current Comet Encke is, however, inadequate to supply the zodiacal cloud, and it is probably just a recently reactivated fragment of the progenitor.

The Taurid Complex has also been detected in situ through the Long Duration Exposure Facility, in the meteoroid mass range $10^{-10}$ to $10^{-4}$~gm, on four large-area faces over a period of 5.8 years (McBride et al. 1995). These workers found that the Taurid Complex is a major part of the current zodiacal cloud, yielding a significant proportion -- say a third -- of the sporadic meteoroid influx. Zook (2001) found that asteroidal grains probably do not contribute more than 25\% of the meteoroid cratering population on LDEF, comet dust being the rest. 
These results appear to strengthen the earlier conclusions of Whipple (1967) and others that the Taurid meteors are the remnants of an erstwhile exceptional body in a short-period, Earth-crossing orbit. 

They are also consistent with the empirical finding that the current zodiacal cloud is overmassive by one or two powers of 10 in relation to current replenishing sources (Whipple 1967; Clube \& Napier 1984; Hughes 1996; Napier 2001). The mass of the zodiacal cloud is estimated to be $\sim$2-30$\times 10^{19}$~gm (Hughes 1996). If say half of this is due to a single progenitor comet, and (conservatively) half its input has disappeared due to Poynting-Robertson and collisional effects, then for progenitor comet density 0.5~gm\,cm$^{-3}$, this  yields an original progenitor comet 50 -- 100~km in diameter, neglecting the contribution from larger bodies within the Complex (Napier 2009).
  
\subsection{Near-Earth objects associated with Comet Encke}

\begin{figure}
\includegraphics[angle=270,width=1.1\linewidth]{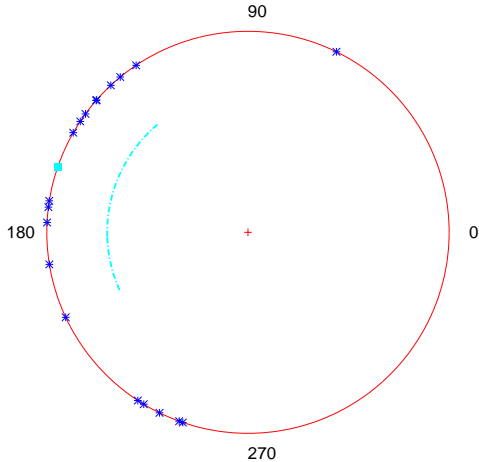}
\caption{Distribution of the longitude of perihelion $\varpi$ of the 19 brightest NEOs ($H<16.5$) with ($a,e,i$) in the range 1.85$\le a \le$2.7, 0.65$\le e\le$1.0 and $0\le i\le 14^\circ$ encompassing the Taurids and Comet Encke, but with no restriction on $\varpi$. These are shown as asterisks. Their mean $\varpi$=174$^\circ \pm12^\circ$, which is 13$^\circ$ from Comet Encke's longitude of perihelion $\varpi\sim \! 161^\circ$, marked with a square. The inner arc gives the spread of the recognized Taurid meteor showers (Stohl 1986). The bottom five asterisks appear to form a separate group, which includes the large near-Earth asteroid Hephaistos.}
\label{perimag}
\end{figure}

\begin{figure}
\includegraphics[angle=270,width=0.9\linewidth]{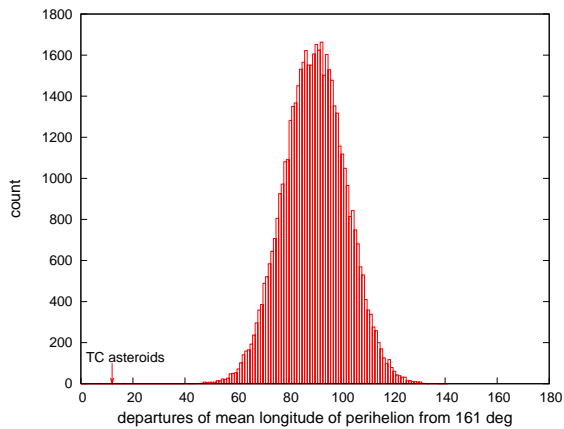}
\caption{Outcome of 50,000 trials in which the longitude of perihelion $\varpi$ of 19 objects is randomly distributed and the mean departure from $\varpi = 161^\circ$ -- that of Comet Encke -- is recorded. For comparison the mean absolute difference of 13$^\circ$ is shown, derived from the 19 NEOs of Fig.~\ref{perimag}. Fragmentation of a large comet is inferred to be the source of these orbital correlations, on a timescale $\sim$20--30~kyr. }
\label{varpi}
\end{figure}

Comet evolution commonly proceeds via fragmentation into smaller bodies, and indeed this may be the prime disintegration route (Jenniskens 2008; Levasseur-Regourd, Hadamcik, Desvoivres et al. 2009). The disintegration of D/1993 F2 Shoemaker-Levy 9 was followed by the impact of $\sim$20 fragments on to Jupiter, with impact energies $\sim$10$^5$~Mt. In late 1995, Comet 73P/Schwassmann-Wachmann~3 split into three fragments, having been seen as a single object on its previous return in 1990. On subsequent returns fresh mini-comets were seen, and by 2006 over 150 fragments had been detected (Reach, Vaubaillon, Kelley et al. 2009; Ishiguro, Usui, Sarugaku et al. 2009). At least one fragment was $\sim$300~m across and a dozen or so were in the size range of the Tunguska impactor. Comet C/1996 B2 Hyakutake split into seven sub-nuclei whose sizes seem to have been in the range of the Tunguska impactor (Desvoivres, Klinger, Levasseur-Regourd et al., 2000); Comet C/1999 S4 LINEAR disintegrated into thousands of fragments. Di Sisto, Fernandez \& Brunini (2009) consider that on average a single Jupiter Family comet ($P<$20~yr) will undergo a major splitting every $\la$77 revolutions. There appears to be more than one mode of disintegration, breakup at perihelion presumably due to tidal stresses being one, and `unknown' being the others.  In the case of the Kreutz sungrazers, several hundred fragments are known of a total probably running to thousands, and it appears that disintegration proceeds through a cascade of fragmentations (Sekanina \& Chodas 2004), the fragments themselves continuing to break up. 

Cometary breakup may also involve the creation of dormant bodies (Kresak \& Kresakova, 1987; Jenniskens  loc. cit., Porub\u{c}an, Korno\u{s} \& Williams 2004; Babadzhanov, Williams \& Kokhirova 2008a). This seems to have happened, for example, in the case of the Quadrantids (Jenniskens 2004; Babadzhanov et al. 2004), and Comet Biela (Jennisken \& Vaubaillon 2007), while fireball outbursts from the $\kappa$~Cygnid shower support the hypothesis that the formation of the stream and dormant bodies within it proceeded from a cascade of disruptions of the progenitor comet (Trigo-Rodriguez et al. 2009).

The presence of substreams in the Taurid Complex indicates that the progenitor comet also took the route of multiple disintegration, and searches have revealed a number of NEOs associated with the sub-streams of the complex. (Clube \& Napier 1984, 1990; Asher, Clube \& Steel 1993; Babadzhanov, Williams \& Kokhirova 2008b, Porub\u{c}an, Korno\u{s} \& Williams 2006). Clube \& Napier (1984) pointed to orbital resemblances between Comet Encke, Oljato, 1982 TA and 1984 KB -- later numbered and named 6063 \nopagebreak{Jason} -- and proposed that these bodies, along with the Taurid meteors, were the remnants of an erstwhile giant comet. These are amongst the largest NEOs. The database has expanded enormously since then, and can be further examined to see whether the hypothesis continues to hold up. The data chosen for this purpose were the NEAs listed in the Minor Planet Center to 2nd April 2008, comprising 2250 Amor (Mars-crossing) asteroids and 2622 Apollos (Earth-crossing).

Fig.~\ref{perimag} shows the result of extracting the 19 brightest near-Earth asteroids ($H<16.5$) in the range 1.83$\le a\le2.7$\,AU, 0.63$\le e\le 1.0$ and $0^\circ\le i\le 14^\circ$ broadly encompassing the elements of Comet Encke, but with no restriction on the longitude of perihelion $\varpi$, which is $\sim$161$^\circ$  for the comet. In the absence of a generic relation between these NEAs, their $\varpi$-distribution is expected to be random, with the longitudes of perihelion precessing at somewhat different rates. However it can be seen that this is not so, the average value of $\varpi$ being $\sim$179$\pm 14^\circ$, i.e. 13$^\circ$ from Comet Encke, as against an expectation of $\sim$90$^\circ$ for a random distribution (Fig.~\ref{varpi}). All but one of these NEOs lie in the 120$^\circ$ range 133$\le\varpi\le 251^\circ$ encompassing Comet Encke. The significance of this result was tested by running trials in which the longitude of perihelion of 19 objects was randomly distributed, and their mean departure from 161$^\circ$ was tested. The outcome of 50,000 such simulations is illustrated in Fig.~\ref{varpi}. Clearly, the configuration shown in Fig.~\ref{perimag}  has negligible probability of being due to chance.

Five of the 19 NEAs in Table~\ref{NEAs} are tightly bunched around the current $\varpi = 237^\circ$ of 2212 Hephaistos, their mean longitude being 238$\pm 7.6^\circ$. They appear to form a separate group (Clube \& Napier 1984; Steel \& Asher 1994). If these and Hephaistos are excluded, the remaining NEAs have
$<\varpi > = 140.8 \pm 10.1^\circ$ as against say $\sim$150.5$^\circ$ for the 15 Taurid substreams identified by Porub\v{c}an, Korno\v{s} \& Williams (2006), or 136.2$^\circ$ for the 10 Taurid Complex asteroids which they identified. The longitudes of perihelion of the Taurid meteors generally lie in the range 100$^\circ < \varpi < 190^\circ$ (e.g. \u{S}tohl 1986).

\begin{table}
\caption{Bright, high eccentricity NEAs selected as described in the text. The longitudes of perihelion tend to cluster in two groups corresponding to the Taurid meteor complex and 2212 Hephaistos respectively.}
\label{NEAs}
\begin{tabular}{lrrrrr} \hline
name            & $H$   &$\varpi$&  $i$   &  $e$   &   $a$  \\ \hline   
1996 FR3        & 16.40 &  63.9  &   8.1 & 0.795 &   2.167 \\
Poseidon        & 15.50 & 123.8  &  11.9 & 0.679 &   1.836 \\
1982 TA         & 14.60 & 129.4  &  12.2 & 0.772 &   2.302 \\
Lugh            & 16.40 & 133.0  &   3.9 & 0.698 &   2.567 \\
Zeus            & 15.80 & 138.8  &  11.4 & 0.654 &   2.262 \\
5025 P-L        & 15.70 & 139.0  &   3.8 & 0.746 &   2.523 \\
1998 SS49       & 15.80 & 143.9  &  10.8 & 0.640 &   1.923 \\
Jason           & 15.30 & 146.5  &   4.9 & 0.763 &   2.220 \\
1999 UM3        & 16.30 & 150.3  &  10.7 & 0.669 &   2.383 \\
Cuno            & 14.40 & 171.0  &   6.8 & 0.636 &   1.981 \\
Oljato          & 15.20 & 172.8  &   2.5 & 0.713 &   2.172 \\
Heracles        & 14.00 & 177.2  &   9.1 & 0.773 &   1.833 \\
1994 AH2        & 16.30 & 189.2  &   9.6 & 0.709 &   2.533 \\
2000 WK63       & 16.20 & 205.0  &  10.4 & 0.758 &   2.438 \\ *[5pt]
Hephaistos      & 13.90 & 236.8  &  11.7 & 0.834 &   2.167 \\
1990 TG1        & 14.70 & 238.8  &   8.7 & 0.680 &   2.440 \\
1990 SM         & 16.10 & 243.9  &  11.6 & 0.766 &   2.101 \\
1999 RD32       & 16.40 & 250.0  &   6.8 & 0.771 &   2.638 \\
Mithra          & 15.60 & 251.1  &   3.0 & 0.661 &   2.204 \\ 
\hline
\end{tabular}
\medskip
\end{table}

A possible objection to the idea of a common origin for these bodies is that, where the surface reflectance properties have been measured, they are more akin to common main belt asteroids than the D-type, carbonaceous surfaces associated with comet nuclei and the more volatile rich, `primitive' regions of the main asteroid belt. While there is spectral diversity amongst the nuclei of comets (Luu 1993), not enough is known about the spectral characteristics of comet nuclei to permit a confident diagnosis about the provenance of any particular small body. Nearly all the NEAs in Table~\ref{NEAs} have perihelia $<$1~\AU, and thermal histories quite different from those of most comets. Sublimation of volatiles at high temperatures from the surface layers may leave behind mantles of rubble relatively depleted in dark organics. A possible example of this is the 5~km object Phaethon, which has the reflectance spectrum of a main belt asteroid, but whose associated Geminid meteors give them an unambiguously cometary origin (Kasuga 2009; Borovi\c{c}ka 2007). Ten of the Taurid Complex asteroids listed in Table~\ref{NEAs} also appear in a group of 17 NEAs obtained by Babadzhanov (2001) using the reduced D-criterion of Asher et al. (1993), when orbital similarity was tested against that of 2P/Encke. The list of Babadzhanov (2001) was not restricted by magnitude. All the NEAs in his list had meteor showers associated with them, supporting a cometary origin for these bodies.

 The alignment shown in Fig.~\ref{perimag} is rapidly lost as one moves to fainter magnitudes. This might imply that, at the sub-kilometre level, there is a dearth of bodies produced by the disintegration of the progenitor comet, or that its signal is lost in the noise of asteroids from the main belt. Detection of such bodies is likely to be difficult, although Porub\u{c}an et al. (2006) and Babadzhanov et al. (2008b) have found dynamical evidence for nine small asteroids, as well as 15 substreams, within the Taurid meteor complex and associated with observable meteor showers and fireballs.
 
In the course of its disintegration, one expects the original comet to have given rise to concentrations of material along its orbital track. This disintegration may be continuing to the present day, but given the evidence that the original comet was exceptionally massive, there may have been epochs of strongly enhanced fireball and Tunguska-like activity during its most active phase. Swarms of particles trapped in resonant orbits would extend the lifetime of some of these concentrations, as predicted by Asher and Clube (1993) and observed by Dubietis \& Arlt (2007). Porub\u{c}an et al. (2006) found that two of the filaments originated 4000 -- 5000 years ago, while one originated over 5000 years ago. The Complex is likely to be much older than this, perhaps 0.2-0.3~Myr (Steel \& Asher 1996). This great age, deduced from the dispersion of the material, is consistent with its exceptional mass. Since the putative 12,900~BP event of North America fits comfortably within the timescale of evolution of the progenitor comet, we may reasonably ask whether one can construct a comet disintegration model to account for the event.
 
\section {Evolution of a large disintegrating comet}

Generally, the separation speeds of cometary fragments are up to $\sim$5~m\,s$^{-1}$ (Kres\'{a}k 1993). However the escape velocity from the surface of a 100~km comet is $\sim$15~m\,s$^{-1}$. As the bodies cascade down through smaller sizes, disruption speeds will presumably decline. Here we consider the case where a comet fragment in an Encke-like orbit disrupts at perihelion. The initial orbit is given an inclination of 6$^\circ$. The fragments are assumed to move away in random directions from the comet. 
\bigskip

\begin{figure}
\includegraphics[angle=270,width=0.9\linewidth]{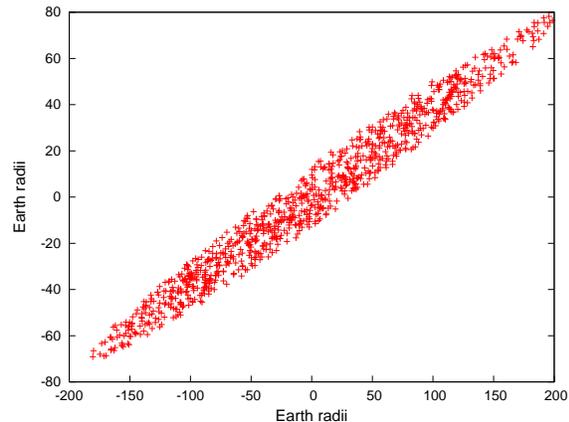}
\caption{A 1,000-particle meteoroid trail viewed from an ecliptic pole. The trail was generated by fragmentating a comet in an Encke-like orbit at perihelion, the fragments being ejected in random directions at speeds up to 20~m\,s$^{-1}$. The Earth would traverse this trail in at most a few hours.}
\label{cod}
\end{figure}
Figure~\ref{cod} shows the projection, looking from an ecliptic pole, of 1,000 particles ejected at perihelion with random speeds up to $V_m$ = 20~m\,s$^{-1}$, during their first passage through a heliocentric sphere of radius 1~\AU. Depending on the encounter geometry, the Earth traverses this trail in at most a few hours. For comparison Kres\'{a}k (1993) found that the duration of rare, intense meteor storms is about an hour. The cross-sectional area of this trail is $\sim$2000 times that of the Earth, but its length increases linearly with time.

The \"{O}pik encounter formula (\"{O}pik 1976) leads to an expectation that, for a single object in an orbit like that of Comet Encke, the mean interval between collisions with the Earth is $\sim$2$\times10^{8}$~yr. For an encounter with a single meteoroid trail with 10$^4$ times the area of the Earth, the interval between such encounters becomes $\sim$2$\times 10^{4}$~yr, comparable with the 20-30~kyr fragmentation history of the Taurid Complex progenitor. Thus if at any given time -- during its major disintegration history -- such a trail is likely to be present, one or more such encounters is a reasonable expectation. If the meteoroid trail comes from the perihelion disintegration of say a $5\times10^{17}$~gm fragment of the original comet, then the Earth would encounter say 1-5$\times 10^{14}$~gm of this material, energetically equivalent to the impact of 2,000-10,000 Tunguska objects over about an hour. 

The fragments from a disrupting comet tend to have size distributions $\propto d^{-q}$ with index $q$ not very different from those expected from self-similar cascades ($q = 3.5$), implying that the mass tends to be concentrated in a few largest bodies (e.g. Ishiguro et al. 2009). The probability of a damaging encounter was investigated by adopting a self-similar fragmentation model, in which half the mass of a disintegrating fragment was assumed to be lost as dust, and the mass of the largest sub-fragment was taken as 0.3 times the initial mass. This yielded five sub-fragments, each of which disintegrated in self similar manner. Assume that the limiting mass for a meteoroid trail -- generated at perihelion and intercepted at Earth -- to cause a damaging bombardment is $m_c =10^{17}$~gm. Each cometary fragment poses a hazard until it disintegrates into sub-fragments and so on, hence the totality of trails created during a cascading disintegration must be computed. With this model we find that a comet whose initial mass is $\sim$10$^{20}$~gm will, in the course of its disintegration, yield $\sim$1,000 trails of mass at least 10$^{17}$~gm. Then by the above argument, if the active lifetime of the comet is say 20,000 years, the swarm must remain as a coherent entity for $\sim$20~yr for an expectation of one such encounter. 

This lifetime requirement is modest, the more so since the orbital period of the Taurid complex bodies ($\sim$3.3~yr) is close to a 7:2 mean motion resonance with Jupiter (3.39~yr), which has a shepherding effect, restricting the spread of dense swarms of meteoroids generated by disintegrations and so prolonging their lifetimes well beyond 20~yr (Asher \& Clube 1993; Beech, Hargrove \& Brown 2004; Dubietis \& Arlt 2007). From a study of historical meteor storms, and comparison with the cometary dust trails detected by IRAS, Kres\'{a}k (1993) found that the survival times of the trails are typically $\sim$60~yr. 

These considerations suggest that, in the course of disintegration of the comet Encke progenitor over 20-30~kyr, one or more damaging encounters with the debris from a recent fragmentation is a reasonably probable event.

\section{Some geophysical expectations}

\begin{table}
\caption{Orbital elements of Lake Tagish bolide (Brown et al. 2000) compared to those of asteroid Cuno (Terentjeva \& Barabanov 2004) and Comet Encke. The elements are for epoch 2000.
}
\label{tagish}
\begin{tabular}{lrrrrr} \hline
object       & $a$ (AU)    &$e$            & $i$         &  $\varpi$   & $T$(yr)  \\ \hline   
Tagish Lake  & 2.1 & 0.57 &  1.4 & 160 & 3.0 \\
(4183) Cuno  & 2.0 & 0.64 &  6.8 & 171 & 2.8 \\         
2P/Encke     & 2.2 & 0.85 & 11.8 & 161 & 3.3  \\
\hline
\end{tabular}
\medskip
\end{table}

We consider here some broadbrush expectations arising from the encounter of a swarm of cometary debris impinging on a hemisphere of the Earth. To match the observations at the Boundary, the swarm model must be able to reproduce the observations of extensive conflagration on at least a continental scale, predict the high concentration of nanodiamonds, and account for the presence of apparently lunar material in the boundary layer.

\subsection{Thermal effects} Wood subjected to a flux of radiant energy $\sim$4$\rm\times 10^7erg\,cm^{-2}\,s^{-1}$ will ignite within 1--20~seconds. While the whole-Earth encounter may take hours, the energy within any local horizon may be released overhead within a few seconds. Wildfires will thus be initiated over a land area say 5\% of the area of the Earth (that of  North America, for example) if the meteoroid swarm deposits $\sim$1 - 20$\times 10^{25}$~ergs during the encounter. At the extreme, we can assume an encounter speed $v$=30~km\,s$^{-1}$ and 100\% efficiency of conversion of kinetic to radiant energy in the upper atmosphere. Then the required mass of swarm to yield a continent-wide conflagration is $\sim$2 - 40$\times 10^{12}$~gm. If the radiant energy is an order of magnitude down on this (say with the most of the energy going into bulk motion of air before disintegration, slowing the bolides to say 10~km\,s$^{-1}$), then one requires the Earth to intercept a mass $\sim$2 - 40$\times 10^{13}$~gm  which is in the range of the masses discussed in Section~3. In reality, this energy will be distributed over a large number of discrete areas corresponding to Tunguska-like fireballs. There is therefore an expectation of conflagration over at least continental scales following such an encounter. A similar figure is reached if one distributes say $N$ epicentres each of 2$\times 10^3$~km$^2$ of fire damage over 25$\times 10^6$~km$^2$ of continent: then one requires $N\sim$10$^4$ Tunguska-like impacts corresponding to an intercepted mass $\sim$5$\times 10^{14}$~gm.

\subsection{Nanodiamonds} The Tagish Lake meteorite fell over Yukon Territory in British Columbia on 18th January 2000. The fragments are extremely fragile, low-density objects with a unique lithology, bearing some resemblance to the most primitive carbonaceous chondrites types, the CI and CM chondrites; the meteorite is nevertheless quite distinct from either of them. It contains unprocessed pre-solar grains and the highest abundance of nanodiamonds of any meteorite, $\sim$3650-4330~ppm (Grady, Verchovsky, Franchi et al. 2002). The pre-entry orbital elements of the body are shown in Table~\ref{tagish}, along with those of asteroid Cuno and Comet Encke. The Tagish Lake fall has been identified with the $\mu$-Orionid fireball stream which, in turn, has been associated with (4183) Cuno (Terentjeva \& Barabanov 2004), an asteroid belonging to the group of large NEOs which appear to be part of the Taurid Complex (Table~1). These generic relationships, along with the primitive nature of the Tagish Lake meteorite, suggest that it may be part of this Complex and that nanodiamonds are generally present in the progenitor comet.

The concentration of nanodiamonds in the 12.9~ka sediments at multiple locations across North America ranges from $\sim$0.01 to 3.7 ppm by weight (Kennett et al. 2009). Let the incoming cosmic material have fractional nanodiamond abundance $f$, falling on an area $A$ of the Earth, and spread through a depth $h$ of the Younger Dryas Boundary. Then to yield a concentration $c$ throughout the boundary, the mass of swarm material intercepted by the Earth is
\begin{equation}
M = A h \rho c/f
\end{equation}
where $\rho$ is the bulk density of the material wherein the nanodiamonds are found. If  the bulk of the swarm material falls on say 5\% of the Earth's surface ($A =2.5 \times 10^{17}$~cm$^2$), then adopting $f = 4 \times 10^{-3}$ by weight from Grady et al. (2002), $c = 10^{-6}$ from Kennett et al. (2009), and assuming the nanodiamonds are concentrated in a layer of thickness $h$ = 2~cm and bulk density $\rho$ = 2.5~gm\,cm$^{-3}$, one finds that intercepting a mass of $6\times 10^{14}$~gm can account for the observed nanodiamond concentration at 12.9~ka. This should probably be increased by a factor of two or so because volatiles in the cometary material would disperse in the atmosphere, but it is still a modest requirement in the context of the breakup model. There is therefore no necessity for the nanodiamonds to be formed in impacts, although of course this is by no means excluded.

\subsection{Lunar debris} The Earth-Moon distance is $\sim$60 Earth radii, comparable with the width of the meteoroid filaments being considered here (Fig.~\ref{cod}); thus any swarm which strikes the Earth may also strike the Moon, the bulk of this impacting mass being in the few largest objects. The mass of lunar material thrown into space by impactors say $\ga 10^{11}$~gm is uncertain but probably comparable to the mass of impacting material (Chyba, Owen \& Ip 1994), and a proportion of this escaped material will strike the Earth. The dynamical evolution of lunar ejecta has been followed by Gladman, Burns, Duncan et al. (1995). The delivery efficiency $p$ depends strongly on the location of impact, but characteristically $p\sim$0.05--0.2 of the lunar material ejected will impact the Earth, the bulk of it within a few years of impact. Thus, if say the mass of the meteoroid swarm material striking the Earth is $M_E$, then because of its smaller target area, the Moon will intercept $0.05 M_E$ of swarm material, resulting in the ejection to space of a similar mass of lunar material and the fall of $0.05 p M_E$ of Moon dust on to the Earth. With $p=0.1$, one expects an admixture of lunar material at the wildfire horizon amounting to $\sim$1\% that of swarm material, to order of magnitude. This seems to be in the ballpark of the KREEP-like material discussed by Firestone (2009).

\section {Summary and conclusions}

The proposition that an exceptionally large comet has been undergoing disintegration in the inner planetary system goes back over 40 years (Whipple 1967), and the evidence for the hypothesis has accumulated to the point where it now seems compelling.
Radio and visual meteor data show that the zodiacal cloud is dominated by a broad stream of largely cometary material which incorporates an ancient, dispersed system of related meteor streams. Embedded within this system are significant numbers of large NEOs, including Comet Encke. Replenishment of the zodiacal cloud is sporadic, with the current cloud being substantially overmassive in relation to current sources. The system is most easily understood as due to the injection and continuing disintegration of a comet 50--100~km in diameter. The fragmentation of comets is now recognized as a major route of their disintegration, and this is consistent with the numerous sub-streams and comoving asteroids observed within the Taurid Complex. The probable epoch of injection of this large comet, $\sim$20--30~kyr ago, comfortably straddles the 12.9~kyr date of the Younger Dryas Boundary.

The hypothesis that terrestrial catastrophes may happen on timescales $\sim$0.1--1~Myr, due to the Earth running through swarms of debris from disintegrating large comets, is likewise not new (Clube \& Napier 1984). However the accumulation of observations has allowed us to build an astronomical model, closely based on the contemporary environment, which can plausibly yield the postulated YDB catastrophe. The interception of $\sim$10$^{15}$~gm of material during the course of the disintegration is shown here to have been a reasonably probable event, capable of yielding destruction on a continental scale.

The object of this paper is not to claim that such an encounter took place at 12,900~BP -- that is a matter for Earth scientists -- but to show that a convincing astronomical scenario can be constructed which seems to give a satisfactory match to the major geophysical features of the Younger Dryas Boundary data. 

If indeed the YDB event was an astronomical catastrophe, its occurrence bears little relation to current impact hazard assessments derived from NEO surveys.

\section*{Acknowledgements} I thank Iwan Williams and David Asher for suggestions which materially improved the original version of the paper.

\end{document}